\documentclass[aps,floatfix,prd,showpacs]{revtex4}
\usepackage{graphicx}% Include figure files
\usepackage{dcolumn}% Align table columns on decimal point
\usepackage{bm}% bold math

\begin{document}

\def\dq{\frac{d^3q}{(2\pi)^3}\,}
\def\be{\begin{equation}}
\def\ee{\end{equation}}

\title{Cusps and Exotic Charmonia}

\author{E.S. Swanson}
\affiliation{
Department of Physics and Astronomy,
University of Pittsburgh,
Pittsburgh, PA 15260,
USA.}

\date{\today}

\begin{abstract}
A simple, causal, and analytic model of final state rescattering is used to describe all available data on the exotic resonances $Z_c(3900)$ and $Z_c(4025)$. The model provides a compelling and accurate representation of experiment with no need for poles in the scattering matrix.
\end{abstract}
\pacs{14.40.Rt, 13.25.Gv}

\maketitle 

\section{Introduction}

The recent confirmation of the $Z(4475)$ resonance by LHCb\cite{Z4475} and the discovery of charged charmonium states $Z_c(3900)$, $Z_c(4025)$ and bottomonium states, $Z_b(10610)$, $Z_b(10650)$\cite{Zb}, point to a possible extensive exotic hadronic spectrum. These discoveries have led to a host of interpretations of these states as loosely bound molecules\cite{mol}, tetraquarks\cite{tet}, hadrocharmonium\cite{hc}, or hybrids\cite{hy}.  This paper explores the simplest possible explanation of the $Z_c(3900)$ and $Z_c(4025)$, namely that the enhancements associated with these states are associated with threshold or cusp effects.

The $Z_c(3900)$ was discovered by the BESIII and Belle collaborations\cite{3900-pipi} in $e^+e^- \to Y(4260) \to J/\psi \pi^+ \pi^-$ in the charged mode $Z_c \to J/\psi \pi^\pm$. The reported mass and width are $M= 3899.0 \pm 3.6 \pm 4.9$ MeV and $\Gamma = 46 \pm 10 \pm 20$ MeV. There is strong evidence that the quantum numbers of the state are $J^P=1^+$. The $Z_c(3900)$ was also observed as a threshold enhancement in the reaction $e^+e^- \to \pi D\bar D^*$\cite{3900-DDst}, where its mass and width were determined to be $3883.9\pm 1.5 \pm 4.2$ MeV and $24.8 \pm 3.3 \pm 11.0$ MeV respectively.

The $Z_c(4025)$ was observed by BESIII in $e^+e^- \to D^* \bar D^* \pi$ at $\sqrt{s} = 4.26$ GeV\cite{4025-DstDst} and in $e^+e^- \to h_c\pi\pi$ at a variety of energies from $\sqrt{s}=  3.90$ to 4.42 GeV\cite{4025-pipi}. 
Its mass was determined to be $4026.3 \pm 2.6 \pm 3.7$  MeV and $4022.9 \pm 0.8 \pm 2.7$  MeV in the respective experiments, while the  measured widths were $24.8 \pm 5.6 \pm 7.7$ MeV and $7.9 \pm 2.7 \pm 2.6$  MeV respectively.

Gaining an understanding of this spectrum is clearly relevant to advancing qualitative (and possibly quantitative) understanding of quantum field theories in their nonperturbative guises. 
In this regard, it is important to develop a sufficiently robust phenomenology to be able to distinguish perturbative and nonperturbative explanations of experimental data. 
For example, it has been noted that cusp effects due to coupled channel thresholds can lead to enhancements in rates that mimic S-matrix poles\cite{bugg,lebed}. This point was recently expanded in Ref. \cite{Zcusp}, where it was shown that a simple model of a four-point function that incorporates a virtual coupled channel ($D\bar D^*$ and  $D^*\bar D^*$ or $B\bar B^*$ and $B^*\bar B^*$) can provide a unified and quantitatively accurate description of the four exotic states $Z_c(3900)$, $Z_c(4025)$, $Z_b(10610)$, and $Z_b(10650)$.

The conclusions of Refs. \cite{bugg,Zcusp} were subsequently challenged by Guo {\it et al.}\cite{Hanhart}, who argued that explaining the processes $Y(4260) \to \pi D\bar D^*$ and $Y(4260) \to \pi\pi J/\psi$ required nonperturbative interactions, which naturally generate S-matrix poles and obviate the cusp explanation of the charged charmonium states. This conclusion was based on a model that assumed pointlike coupling between relevant hadrons ($\pi$, $Y(4260)$, $D$, $D^*$, and $J/\psi$). The threshold enhancement in $Y \to \pi D\bar D^*$ was then generated via a loop with an intermediate $D\bar D^*$ state that was regulated with a phenomenological exponential cutoff function. The resulting coupling constants were too small to provide agreement with $Y\to \pi\pi J/\psi$ data unless the loop diagrams were iterated, which led the authors to conclude that the $Z_c$ enhancements are true resonances.

The purpose of this paper is to examine the robustness of the conclusions of Guo \emph{et al.} by constructing an explicit model for the interactions of the relevant hadrons. This will also permit checking the conclusions of the model presented in Ref. \cite{Zcusp}. The proposed model is both simple and plausible. Furthermore, its predictions follow from the Schr\"{o}dinger equation, thereby avoiding the criticism of Ref. \cite{as}. Model predictions are in good agreement with the available charmonium data and provide strong evidence that dynamically generated poles are \emph{not} necessary to explain the novel charged charmonium states.

\section{Coupled Channel Model}

The proposed model will employ nonrelativistic dynamics and separable hadronic vertices.
It is a simple matter to relax these choices; however, doing so will not affect results in a qualitative way, and we find it preferable to compute with the simplest model possible to illustrate the relevant effects.  The model vertices are of the form

\be
AB : CD \to  g_{AB:CD} \cdot k_{AB}^{\ell_{AB}}\,\exp(-k^2_{AB}/\beta^2_{AB}) \cdot k_{CD}^{\ell_{CD}}\,\exp(-k^2_{CD}/\beta^2_{CD}),
\ee
where $g$ is the coupling for the relevant channel, $\beta$ is a hadronic scale of order $\Lambda_{\rm QCD}$, and $\ell$ is an integer specifying an angular momentum factor. With the exception of $\ell_{h_c \pi} =1$, all other channels are dominated by S-wave scattering, hence $\ell=0$. The momenta, $k$, are given by

%\begin{eqnarray}
%Y_i \pi :  D D^*_i &\to& g_{YD} \exp(-\lambda_{Y\pi}) \exp(-\lambda_{DD^*}) \nonumber \\
%Y_i \pi  : D^*_j D^*_\ell \epsilon^{ij\ell} &\to& g_{YD^*}  \exp(-\lambda_{Y\pi}) \exp(-\lambda_{D^*D^*}) \nonumber \\
%\psi_i \pi :  D D^*_i &\to& g_{\psi D} \exp(-\lambda_{\psi\pi}) \exp(-\lambda_{DD^*}) \nonumber \\
%\psi_i \pi :  D^*_j D^*_\ell \epsilon^{ij\ell} &\to& g_{\psi D^*}  \exp(-\lambda_{\psi\pi}) \exp(-\lambda_{D^*D^*}) \nonumber \\
%h_c^i \pi :  D D^*_j k_\ell \epsilon_{ij\ell} &\to& k g_{hD} \exp(-\lambda_{h_c\pi}) \exp(-\lambda_{DD^*}) \nonumber \\
%k_i h_c^i \pi : D^*_j D^*_j &\to& k g_{hD^*} \exp(-\lambda_{h_c\pi}) \exp(-\lambda_{D^*D^*}) .
%\label{v1}
%\end{eqnarray}
%ratio of partial widths is DD*/pi psi = 6.2 (1.1) (2.7)

\be
k_{AB}^2 = \lambda(s_{AB}; m_A^2, m_B^2)/(4 s_{AB})
\label{v1}
\ee
where $\lambda$ is the K\"{a}ll\'{e}n function and  $s_{AB} = (p_A+p_B)^2$. This model will be used for the channels $Y \pi : D \bar D^*$, $Y \pi : D^* \bar D^*$, $J/\psi \pi : D\bar D^*$, $J/\psi \pi : D^* \bar D^*$, $h_c \pi : D\bar D^*$, and $h_c \pi : D^*\bar D^*$.   Appropriate charge conjugation is included where required. We remark that these vertices are often represented as $t$-channel $D$ or  $D^*$ exchange diagrams.

\begin{figure}[ht]
\includegraphics[width=10cm,angle=0]{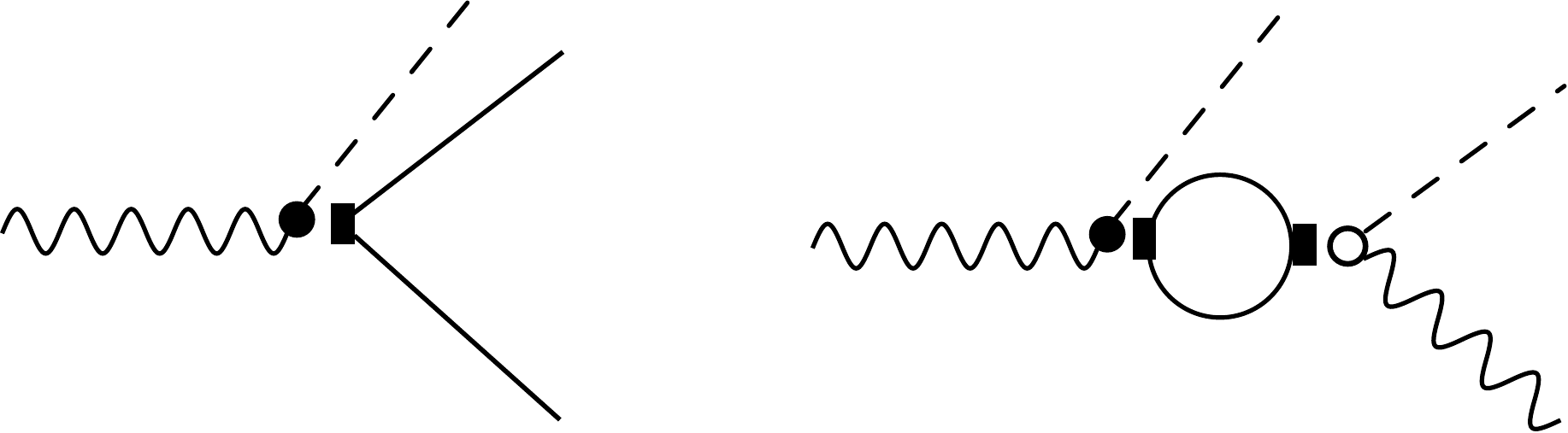}
\caption{Leading order diagrams for $Y \to  \pi D^* \bar D (\bar D^*)$ (left) and $Y \to \pi \pi J/\psi(h_c)$ (right).}
\label{fig-vertex}
\end{figure}

Rescattering will be accommodated with similar vertices for the channels $D \bar D^* : D \bar D^*$ and $D^* \bar D^* : D^* \bar D^*$. Cross channel scattering $D\bar D^* \to D^* \bar D^*$ is not considered in the following.  In other models, these vertices are commonly described by pion-exchange diagrams.  It will be assumed that a direct coupling, $Y\pi : J/\psi \pi$, is negligible; hence this process will be generated by iterations of diagrams of the type shown in Fig. \ref{fig-vertex}(right).

Exponential form factors are adopted in the model vertices because hadronic constituents are not resolved at the low energies being considered here. We regard this assumption as more appropriate for the study of $Z_c$ production than constant couplings.
In general, each of the exponentials can assume a different scale, $\beta_{AB}$. However, for simplicity we shall assume that the $\beta$s take on a single value for almost all of the form factors (the exception will be that for the $DD^*$ form factor, as described below).

Fig. \ref{fig-vertex} shows the leading order diagrams for the processes considered here. All of the results presented below will be unitarized by summing the bubble diagrams that contribute in $s$-channel. The effect of this unitarization can be strong: sufficiently attractive couplings lead to a 
$D \bar D^*$ or $D^*\bar D^*$ bound state and give a sharp peak below threshold in the appropriate channel. Similarly, a repulsive coupling smears out the threshold cusp and pushes it to higher invariant mass. It is thus possible to determine the size of these effects by comparison to experiment.

\section{Charmonium Cusp States}

The model scales $\beta_{Y\pi}$, $\beta_{DD^*}$, $\beta_{D^*D^*}$  and the couplings $g_{Y\pi:DD^*}$, $g_{Y\pi:D^*D^*}$, $g_{DD^*:DD^*}$, and $g_{D^*D^*:D^*D^*}$ will be fixed by comparison to the threshold behavior of the processes $Y \to \pi D\bar D^*$ and $Y \to \pi D^*\bar D^*$. 
Since the purpose of this computation is not a detailed analysis  of experimental results, a fit to the Dalitz plot densities will not be attempted, except where required.

\subsection{$Y \to \pi D^*\bar D^*$}

\begin{figure}[ht]
\includegraphics[width=10cm,angle=0]{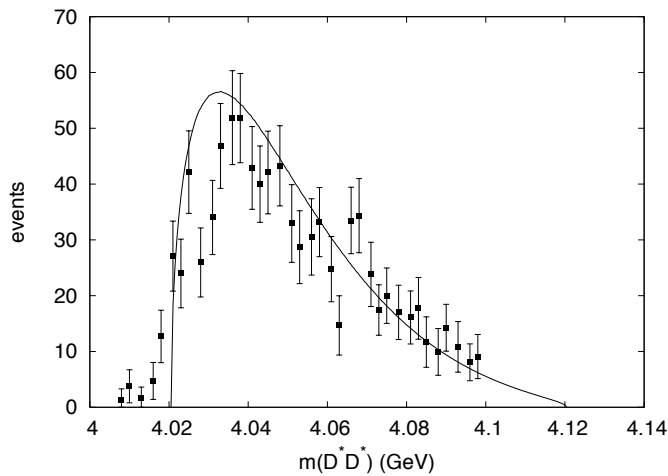}
\caption{\protect $e^+e^-(\sqrt{s}=4.26\ {\rm GeV}) \to \pi D^* \bar D^*$. Solid line: model fit. Data from Ref. \cite{4025-DstDst}.}
\label{fig-DstDst}
\end{figure}

Fig. \ref{fig-DstDst} displays the projection of the $Y\to \pi D^* \bar D^*$ Dalitz plot onto the $m(D^*\bar D^*)$ axis. 
The overall scale is set by the coupling $g_{Y\pi:D^*D^*}$, which is arbitrary since cross sections were not measured.  The solid line shows the result of a fit that evidently describes the data quite well.  The scales chosen were $\beta_{Y\pi}$ = $\beta_{D^*D^*}$ = 0.3 GeV. This value is strongly selected by the data since deviations lead to quite different shapes. Similarly, $D^* \bar D^*$ rescattering also affects the predicted rate and the result

\be
g_{D^*D^*:D^*D^*} \approx 0
\ee
is strongly preferred. Thus the reaction $Y\to \pi D^* \bar D^*$ implies that \emph{no dynamical  $D^*\bar D^*$ resonances will be generated} in any of the following work. We also observe that there is no need for a constant (coupling)  term in the amplitude, in opposition to what was assumed in Ref. \cite{Hanhart} for $Y \to \pi D\bar D^*$.

\subsection{$Y \to \pi D \bar D^*$}

\begin{figure}[ht]
\includegraphics[width=10cm,angle=0]{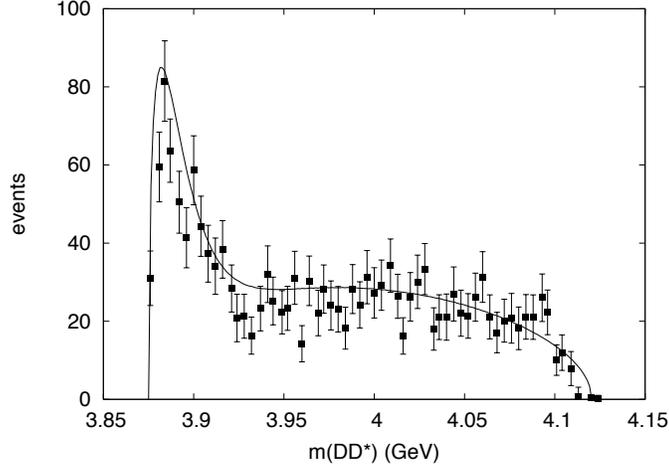}
\caption{$e^+e^-(\sqrt{s}=4.26\ {\rm GeV}) \to \pi D \bar D^*$. Solid line: $\beta_{DD^*} = 0.2$ GeV. Data from Ref. \cite{3900-DDst}.}
\label{fig-DDst}
\end{figure}

Data for $Y \to \pi D \bar D^*$ are presented in Fig. \ref{fig-DDst}.  In attempting to fit this, we note that only a pion and a single $D$ meson were reconstructed during data-taking -- other particles were inferred\cite{3900-pipi}. Thus it is likely that an incoherent background exists in this data and we therefore choose to model this reaction by incorporating such a background as a constant.   The scale $\beta_{Y\pi}$ has already been fixed to 0.3 GeV; this leaves $\beta_{DD^*}$ and the coupling $g_{Y\pi:DD^*}$ (which sets the normalization) and $g_{DD^*:DD^*}$ to determine from the data.

A rough fit (no systematic fitting was made in this work as it was thought to be antithetical to the approach) yielded $\beta_{DD^*} = 0.2$ GeV (shown as a solid line in Fig. \ref{fig-DDst}). The fit is gratifyingly faithful to the data.

%incoherent bg=3.5d-3; bg=0

Once again, the data present no evidence for strong rescattering, and we determine $g_{DD^*:DD^*} \approx 0$, implying that \emph{no $D\bar D^*$ resonances can be dynamically generated}.   These conclusions disagree with those of Guo \emph{et al.}, who assume that a constant $g_{Y\pi: DD^*}$ vertex gives rise to the events seen at large $DD^*$ invariant mass. As we have seen, this is at odds with the data for $Y \to \pi D^*\bar D^*$ and is not necessary to describe $Y\to \pi D \bar D^*$. Furthermore, preliminary results from BESIII do indeed indicate that the background is strongly suppressed when all three final particles are reconstructed\cite{gradl}, providing further support for the model presented here.

\subsection{$Y \to \pi^+\pi^- J/\psi$}

\begin{figure}[ht]
\includegraphics[width=8cm,angle=0]{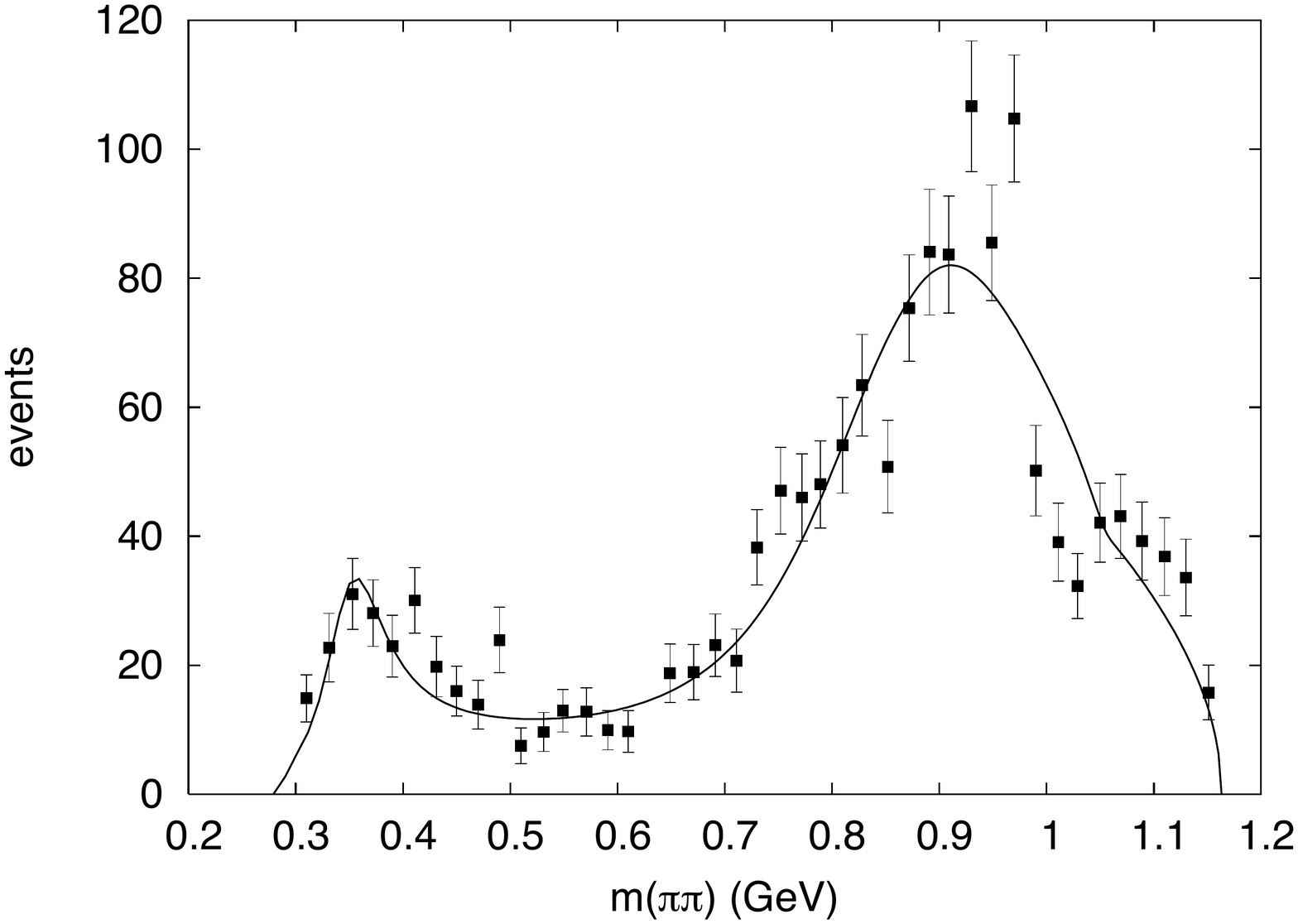}
\qquad
\includegraphics[width=8cm,angle=0]{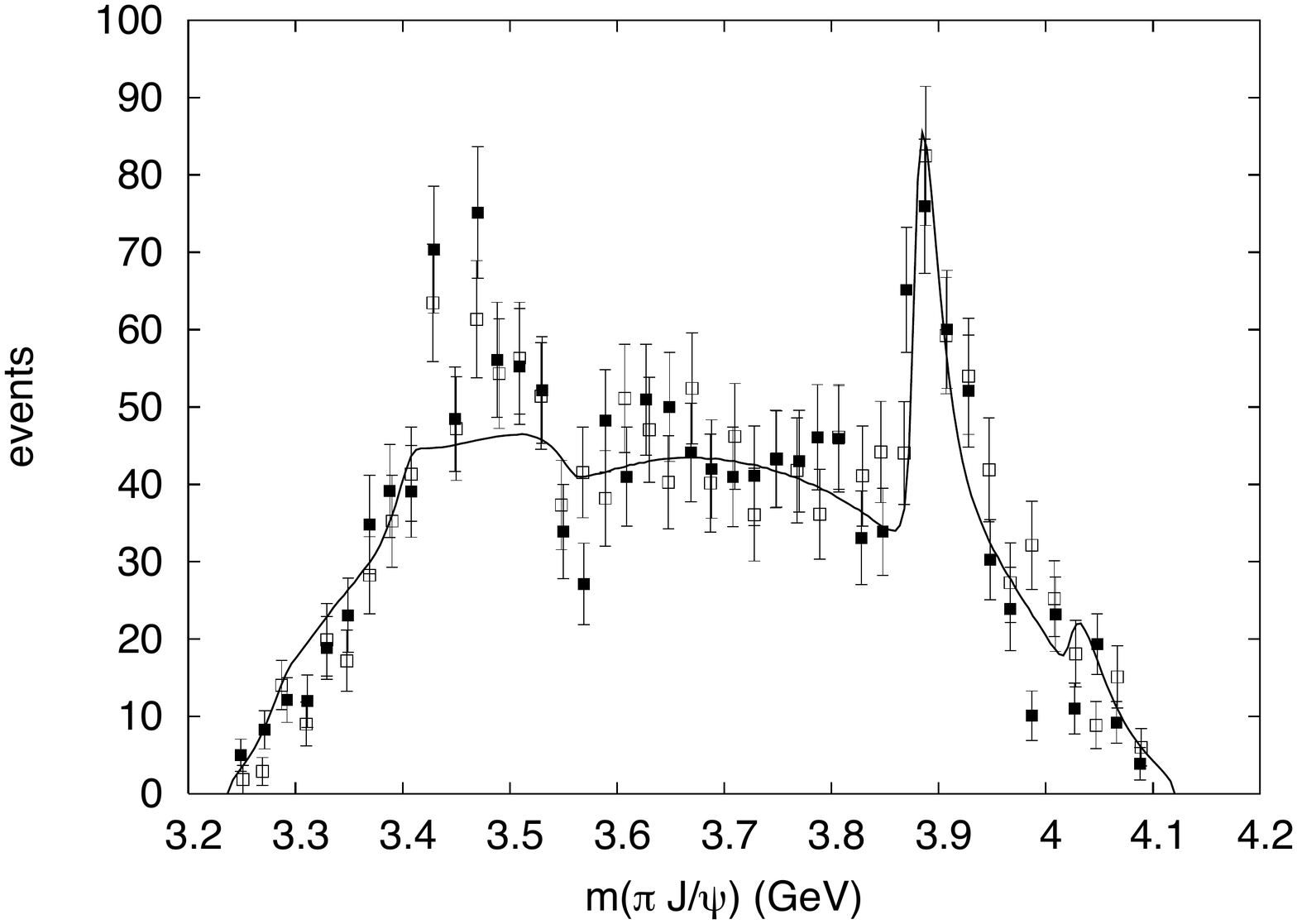}
\caption{$e^+e^-(\sqrt{s} = 4.26\ {\rm GeV})  \to \pi \pi J/\psi$. Left panel: invariant $\pi\pi$ mass distribution. Right panel: invariant $\pi J/\psi$ mass distribution. Filled squares: $\pi^- J/\psi$; open squares: $\pi^+ J/\psi$. Data from Ref. \cite{3900-pipi}.}
\label{fig-pipsi}
\end{figure}

With all the relevant scales and the $D^{(*)}\bar D^*$ couplings fixed, it is possible to make a prediction for $Y \to \pi\pi J/\psi$. Unfortunately, this process is complicated by strong final state interactions in the $\pi\pi$ subsystem, shown in the left panel of Fig. \ref{fig-pipsi}.  Thus a reasonably faithful reproduction of the data requires incorporating $\pi\pi$ dynamics in the amplitude model. Because pion dynamics is not the thrust of the current investigation, a pair of Breit-Wigner amplitudes at 0.35 GeV and 0.88 GeV with widths of 68 and 290 MeV respectively were used to obtain an approximate fit to the $\pi\pi$ spectrum (shown as a solid line in Fig. \ref{fig-pipsi}). A Flatt\'{e} parameterization of the $f_0(980)$ was also attempted but this did not fit the data well.

A constant coherent background is assumed and the assumption $g_{J/\psi \pi: DD^*}$ = $2 g_{J/\psi \pi: D^*D^*}$ is made. 
With the model of $\pi\pi$ dynamics in place, the amplitude is fixed (including the overall normalization, which is determined by the $\pi\pi$ distribution). The resulting prediction is shown as a solid line in the right panel of Fig. \ref{fig-pipsi}. The overall quality of the prediction is quite good, although some strength is missing through the $Z_c(3900)$ reflection near 3450 MeV.  Note that there is a slight enhancement in the prediction that corresponds to the $Z_c(4025)$ cusp in $D^*\bar D^*$. The relative size of this enhancement is controlled by the ratio $g_{J/\psi \pi: DD^*}/g_{J/\psi \pi: D^*D^*}$. This is expected to be of order unity. As mentioned above, the ratio has been set to 2, but in principle could be larger, so that the $D^*\bar D^*$ cusp would be barely visible in the plot.

\subsection{$e^+e^- \to \pi^+\pi^-h_c$}

\begin{figure}[ht]
\includegraphics[width=10cm,angle=0]{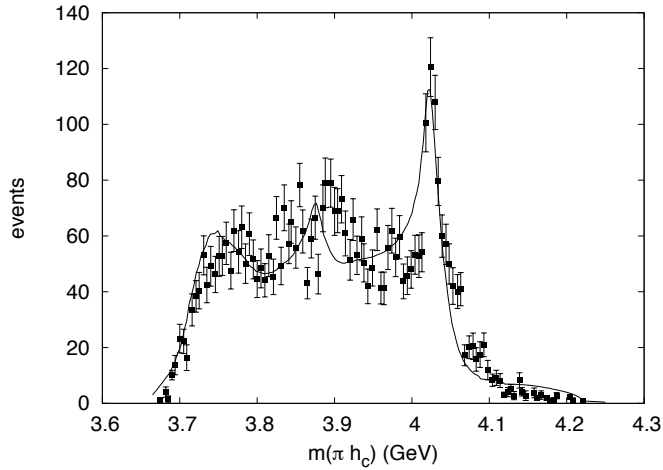}
\caption{$e^+e^- \to \pi^+\pi^- h_c$. Solid line: model fit. Data from Ref. \cite{4025-pipi}.}
\label{fig-Zc4020}
\end{figure}

The $Z_c(4025)$ was observed in $e^+e^- \to \pi\pi h_c$, which was measured by BESIII at 13 values of $\sqrt{s}$. The total event rate summed over all energies is reproduced as data points in Fig. \ref{fig-Zc4020}. 
The data were modelled by generating 13 Dalitz plots corresponding to the experimental values of $\sqrt{s}$ and summing these with a weight given by the reconstructed number of $h_c$ mesons for each energy. All couplings and scales are fixed as above, with the exception of $\beta_{h_c \pi}$, which was set to 0.3 GeV for simplicity.

%coherent bg = 6d-4, phase=pi, coupling for cusps =1, factor=300000
% pipi does not do much (but this depends on couplings...)

In principle, $\pi\pi$ rescattering should affect the predicted distribution. However, we have found that employing the fit dynamics from $Y\to \pi\pi J/\psi$ does not change the results much; thus a simple constant coherent background  was employed in the fit shown here.  Finally, the relative couplings were fit to the data to obtain $g_{h_c \pi:D^*D^*} \approx 2 g_{h_c\pi:DD^*}$. Note that the analogous ratio is approximately inverted in the case of the $J/\psi \pi\pi$ final state.

Once again, the agreement with data is quite satisfactory, especially considering the simplicity of the model. The peak near 4.0 GeV is a cusp effect due to $D^*\bar D^*$ rescattering; the peak near 3.9 GeV is due to $D\bar D^*$ rescattering, and the peak near 3.75 GeV is a reflection of the $Z_c(4025)$ cusp. None of these features are associated with poles in the S-matrix.

\subsection{$\mu N \to \mu\, J/\psi\, \pi^\pm \, N$}

Lin \emph{et al.} have suggested that the coupling of the $Z_c(3900)$ to $\pi J/\psi$ can be exploited to search for it in photoproduction\cite{lin}. The idea is that the virtual photon converts to a $J/\psi$ via the vector meson dominance mechanism, which interacts with a nucleon by pion exchange, creates an $s$-channel $Z_c$, which finally decays to $\pi J/\psi$.  The cross section for $\gamma N \to Z N$ was estimated using a hadronic Lagrangian with dipole form factors. The $Z_c \pi J/\psi$ coupling was taken from the measured width of the $Z_c$ (assuming that it is saturated by this mode).
The resulting cross section was predicted to peak at $\sqrt{s} \approx 7$  GeV with a readily observable rate.

In spite of these expectations, a measurement of $\mu N \to \mu J/\psi \pi N$ by the COMPASS collaboration\cite{compass} found no evidence for the $Z_c(3900)$.
This lack of evidence has a simple explanation in the present model. Dipole form factors should be replaced with the form factors of Eq. \ref{v1}, which are heavily suppressed by the large center of mass energy of the process.
Of course, this observation has no bearing on whether the $Z_c$ is a dynamically generated resonance  since the coupling to a resonance state could be similarly suppressed. Furthermore, the model vertices should not be trusted at very large center of mass energies since this is where the pomeron trajectory is expected to take over the dynamics.

\section{Conclusions}

A detailed comparison to available $Z_c$ experimental data has been presented. The model successfully captures the features of all the data and indicates that there is no evidence for strong  $D\bar D^*$ or $D^*\bar D^*$ rescattering in this system. We note that this conclusion is supported by recent lattice gauge computations that report only weakly repulsive $(D^*\bar D^*)^\pm$ interactions in the $J^P=1^+$ channel\cite{chen}. Thus isovector rescattering is not sufficiently attractive to generate dynamical bound states (this is also supported by lattice computations\cite{Zc-lattice}) and exotic resonances are not required to explain the data.  However, cusp effects remain and can provide a qualitative, and even quantitative, explanation of all of the data.

This simple model is fully quantum mechanical and therefore removes concerns about causality\cite{as}. Furthermore, it appears to provide a better, and more complete, description of the available data than that of Ref. \cite{Hanhart}.

Threshold enhancements and openings are generic features in hadronic systems and one must therefore be cautious in claiming bound states where such effects are known to operate. Certainly, any near-threshold enhancement can simply arise because hadrons are soft; thus the $D\bar D^*$ and $D^*\bar D^*$ data are easily explained. Similarly, coupled-channel cusps should be regarded as the possible explanation for bumps seen in rescattering channels slightly above coupled channel thresholds. Furthermore, if the ``widths" of these enhancements vary strongly (as they do for the $Z_c$s) between threshold and rescattering processes, then this is an additional sign that non-resonant explanations  must be considered. In particular, threshold bumps arise due to competing effects between form factors and phase space, whereas a rescattering enhancement width is mediated by form factors and a rescattering loop. Importantly, this implies that (cusp dominated) threshold bumps do not exhibit phase motion, while rescattering enhancements may have phase motion due to the associated bubble diagrams.

\acknowledgments

I am grateful to Wolfgang Gradl, Christoph Hanhart, Ryan Mitchell, Matt Shepherd, Adam Szczepaniak, and Qiang Zhao for discussions on this topic.

\end{document}